\documentstyle[preprint,aps]{revtex}

\begin{document}
\title{The Critical Properties of a Modulated Quantum Sine-Gordon Model }
\author{Zhiguo Wang, Yumei Zhang}
\address{Department of Physics, Tongji University, Shanghai 200092, China PR}
\date{\today }
\maketitle
\pacs{05.70.Jk, 68.10.-m,87.22.Bt}

\begin{abstract}
A new procedure of trial variational wave functional is proposed for
investigating the mass renormailzation and the local structure of the ground
state of a one-dimensional quantum sine-Gordon model with linear spatial
modulation, whose ground state differs from that without modulation. The
phase diagram obtained in parameters $(\alpha \Lambda ^{-2},\beta ^{2})$
plane shows that the vertical part of the boundary between soliton lattice
phase and incommensurate (IC) phase with vanishing gap sticks at $\beta ^{2}$
$=4\pi $, the IC phase can only appear for $\beta ^{2}$ $\geq 4\pi $ and the
IC phase regime is enlarged with increasing spatial modulation in the case
of definite parameter $\alpha \Lambda ^{-2}$. The transition is of the
continuous type on the vertical part of the boundary, while it is of the
first order on the boundary for $\beta ^{2}>4\pi $.

PACS: 05.70.Jk, 68.10.-m,87.22.Bt

Keywords: Sine-Gordon model, \ spatial modulation, soliton lattice, \
incommensurate phase, \ phase diagram
\end{abstract}

\section{\bf Introduction}

Recently, many experiments$^{\cite%
{Fagot-Revurat,Kiryukhin1,Kiryukhin2,Oshikawa,Wang,Swanson,Koppen,Horvatic}}$
have detected there exists a soliton lattice phase between spatially
homogenous commensurate phase with finite gap and incommensurate (IC) phase
with vanishing gap (free massless boson field) when some
quasi-one-dimensional(1D) spin-Peierls (SP) materials were placed in an
external magnetic field. These experiments confirm the 20-year-old
prediction of a field-induced transition from the SP state into a
spin-lattice modulated phase$^{\cite{Kiryukhin1}}$ and demonstrated that
this novel phase has the form of a soliton lattice.$^{\cite{Kiryukhin2}}$
The density of solitons can be tuned continuously by adjusting the magnetic
field, so that their internal structure can be systematically studied.

In deed, physical realization of SP systems offers the unique possibility to
check the understanding of this fundamental 1D Hamiltonian for arbitrary
band filling. Applying the canonical Jordon-Wigner transformation, the
Hamiltonian of such a SP chain placed in an external magnetic field $H$ can
be mapped onto that of interacting (spinless) fermions system with band
filling tuned by $H$. An convenient way to deal with 1D fermi systems is
bosonization. After adopting the bosonization technique, the Hamiltonian of
such a 1D fermi system is transformed into a modified sine-Gordon model with
a linear spatial modulation in the cosine potential. In the boson picture,
the effective Hamiltonian is

\begin{equation}
H=\int dx\left\{ \frac{1}{2}\left[ \Pi ^{2}(x)+\left( \partial _{x}\phi
\right) ^{2}\right] -\frac{\alpha }{\beta ^{2}}\cos \left( \beta \phi
+\lambda x\right) \right\} \text{ .}
\end{equation}
Here $\phi (x)$ is a boson field operator, $\Pi (x)$ is its conjugate
momentum, $\Pi (x)=-i\frac{\delta }{\delta \phi (x)}$, and they satisfy the
commutation relation 
\begin{equation}
\left[ \phi (x),\Pi (y)\right] =i\delta (x-y)\text{ . }
\end{equation}
$\lambda $ is a spatial modulation parameter, which represents the effect of
fermi surface shifting from half filling in the fermion picture.$^{\text{%
\cite{Sun}}}$

Although the Zeeman energy influences the comparison between the ground
state energy of boson models with different modulation, it has no effect to
the structure of the ground state for fixed band filling case. It is
neglected in present paper.

An alternative realization of the Hamiltonian (1) with fixed $\lambda $
refers to the interaction fermion system with tunable particle
concentration. For example, the charge-density-wave degree freedom of the 1D
extended Hubbard model$^{\cite{Voit}}$ is described by this model. Away from
the half filling, or $\lambda \neq 0$, people usually assumes that the
cosine potential can be neglected due to the rapid oscillation.$^{\cite{Penc}%
}$ However, according to this statement, the energy gap would drop suddenly
to zero at $\lambda =0$. We believe it is more likely that the cosine
potential might experience an intermediate lattice phase before entering
into the free boson field regime.

In absence of $\lambda $, the Hamiltonian is a standard quantum sine-Gordon
model. Many works have been done about this case both in field theory and in
condensed matter physics.$^{\cite%
{Sklyanin,Coleman,Ingermanson,Mandelstam,Stevenson}}$ By a variational
method Coleman$^{\text{\cite{Coleman}}}$ first discovered that the energy
density of the system is unbounded below when the coupling constant $\beta
^{2}$ exceeds $8\pi $, this corresponds to the Kosterlitz-Thouless (K-T)
phase transition by the equivalence between the 2D Coulomb gas and
sine-Gordon model, the ground state is a spatially homogenous commensurate
phase.

In presence of finite $\lambda $, the spatial modulation parameter favors
the IC phase whereas the cosine term itself favors the commensurate phase,
this makes the model more complicated. Previously, many results for the
Hamiltonian (1) have been obtained in the classical limit.$^{\cite%
{Fetter,Sutherland,Fukuyama}}$ In the quantum case, this cosine potential is
usually considered to shrink to zero with any finite $\lambda $, the
crossover between commensurate and IC phases would happen abruptly. Schulz
has discussed a similar quantum system as Hamiltonian (1)$^{\text{\cite%
{Schulz}}}$, he found that the domain wall structure appeared in the ground
state in the case of the chemical potential exceeding a threshold value and
the energy gap was finite, for large chemical potential the presence of the
gap was unimportant. In fact, his result implied that there existed a
soliton lattice phase which had finite density of domain walls between
commensurate phase with finite gap and IC phase with vanishing gap.

After taking a shift $\phi (x)+\frac{\lambda x}{\beta }\rightarrow \phi (x)$%
, the Hamiltonian (1) seems as same as that of Schulz. But if one pays
attention to the boundary condition of boson operator $\phi (x)$, he will
find that two models are different since the boundary conditions of system
will be altered under this shift, namely, $\int_{0}^{L}\partial _{x}\phi
(x)dx=0\rightarrow \int_{0}^{L}\partial _{x}\phi (x)dx=\frac{\lambda }{\beta 
}L$, therefore the physical properties will be also changed. We prefer the
former (zero charge sector), since this integral is proporational to the
particle number after normal ordering in the fermion picture. To our
knowledge, with finite $\lambda $, the complicated configuration between
commensurate and IC phases has not been explored clearly.

In present paper, we develop a new method to give a quantitative discussion
on the soliton lattice phase of the Hamiltonian (1). Our basic idea refers
to a variational procedure with respect to a spatially inhomogenious
Gaussian wave functional which is introduced to determine an approximate
ground state of the model. Our results show that as long as $\lambda $
shifts from zero, the soliton lattice phase appears in the ground state for $%
\beta ^{2}<4\pi $. Only in the regime of $\beta ^{2}>4\pi $ the IC phase
with vanishing gap can emerge in the ground state for small $\alpha $. The
paper is set up as following. In section II, we introduce a new spatially
inhomogenious Gaussian wave functional and calculate the energy and the
renormalization mass of the ground state. The phase diagram and the
evolvement of the soliton lattice phase with varying spatial modulation
parameter are given in section III. Section IV is our conclusion.

\section{Spatially inhomogenious Gaussian wave functional method}

In the Schrodinger field picture, the problem is to solve the functional
Schrodinger equation 
\begin{equation}
H\Psi =E\Psi
\end{equation}
where $\Psi $ is a wave functional of the variable $\phi (x)$. It is
obviously difficult and far beyond our present abilities to solve this
equation for the Hamiltonian (1). Here we try a variational method by
choosing some general ansatz for $\Psi $, then we calculate the expectation
value of the Hamiltonian and minimize the energy with respect to the
variational parameters in the ansatz.

If we consider the field operator $\phi (x)=0$, the cosine term will
oscillate in the space yielding zero contribution to the ground state. This
may not be the most favorable configuration of the ground states. One may
shift $\phi (x)$ to compensate $\lambda x$ as much as possible to lower the
value of the cosine potential term, however, the integration of $\partial
_{x}\phi $ should be zero since it must satisfy the boundary condition. Thus 
$\phi (x)$ should take a kink after some distance, moreover, the modulation
of $\phi (x)$ induces a ''distortion energy'' $\frac{1}{2}$($\partial
_{x}\phi )^{2}$. Nevertheless, we expect an appropriate form of $\phi (x)$
which can lower the total energy at most.

Taking above into consideration, we approach the ground state of the system
by a trial Gaussian wave functional with a field shift as 
\begin{equation}
\Psi (P,\phi _{cl},f)=N_{f}exp\left\{ i\int P(x)\phi (x)dx-\frac{1}{2}\int
dxdy\left[ \left( \phi (x)-\phi _{cl}(x)\right) f(x,y)\left( \phi (y)-\phi
_{cl}(y)\right) \right] \right\} .
\end{equation}
Where $N_{f}$ is the normalization factor, $P(x)$, $\phi _{cl}(x)$ and $%
f(x,y)$ are the variational functions. The expectation value of the
Hamiltonian of Eq.(1) with respect to the wave functional of Eq.(4) is given
in as 
\begin{eqnarray}
E(P,\phi _{cl},f) &=&%
\displaystyle\int 
dx\left\{ \frac{1}{2}\left[ P(x)^{2}+\left( \partial _{x}\phi
_{cl}(x)\right) ^{2}\right] -\frac{\alpha }{\beta ^{2}}Z\cos \left( \beta
\phi _{cl}(x)+\lambda x\right) \right.  \nonumber \\
&&\left. +\frac{1}{4}f(x,x)-\frac{1}{4}\int dy\delta (x-y)\frac{\partial ^{2}%
}{\partial x\partial y}f^{-1}(x,y)\right\} \text{ ,}
\end{eqnarray}
where 
\begin{equation}
Z=\exp \left\{ -\frac{\beta ^{2}}{4}f^{-1}(x,x)\right\} \text{ ,}
\label{eq3}
\end{equation}
$f^{-1}(x,y)$ denotes the inverse of $f(x,y)$, i.e., 
\[
\int f(x,x^{\prime })f^{-1}(x^{\prime },y)dx^{\prime }=\delta (x-y)\text{ }. 
\]
For simplicity we choose $f(x,y)$ of the form 
\[
f(x,y)=\frac{1}{2\pi }\int dk\sqrt{k^{2}+\mu ^{2}}\cos k(x-y)\text{ } 
\]
with inverse 
\begin{equation}
f^{-1}(x,y)=\frac{1}{2\pi }\int dk\frac{\cos k(x-y)}{\sqrt{k^{2}+\mu ^{2}}}%
\text{ ,}  \label{eq2}
\end{equation}
where $\mu ^{2}$ is a variational parameter.

Up to now, we have introduced three variational parameters $P(x)$, $\phi
_{cl}(x)$ and $\mu ^{2}$ which will be determined in the following
calculation. For the ground state, the minimal energy configuration is
clearly achieved with $P(x)=0$.

Introducing a upper cutoff $\Lambda $ in the integral of Eq.(7), Eq.(6) is
explicitly evaluated as 
\begin{equation}
Z=\left( \frac{\mu \Lambda ^{-1}}{1+\sqrt{1+\left( \mu \Lambda ^{-1}\right)
^{2}}}\right) ^{\frac{\beta ^{2}}{4\pi }}\text{ ,}
\end{equation}
and the expectation energy is 
\begin{equation}
E(\phi _{cl},\mu ^{2})=%
\displaystyle\int 
dx\left\{ \frac{\Lambda ^{2}\sqrt{1+(\mu \Lambda ^{-1})^{2}}}{4\pi }+\frac{1%
}{2}\left( \partial _{x}\phi _{cl}(x)\right) ^{2}-\frac{\alpha }{\beta ^{2}}%
Z\cos \left( \beta \phi _{cl}(x)+\lambda x\right) \right\}
\end{equation}

After minimizing the energy expectation value with respect to $\mu $ and $%
\phi _{cl}(x)$ respectively, we get 
\[
\mu ^{2}=\alpha Z\frac{1}{L}\int dx\cos \theta (x)\text{ ,} 
\]
\begin{equation}
\frac{\partial ^{2}\theta (x)}{\partial x^{2}}=\alpha Z\sin \theta (x)\text{
.}
\end{equation}
Here, we have set $\theta (x)=\beta \phi _{cl}(x)+\lambda x$ for brevity, $L$
is the macroscopic length of the system.

Eq.(10) is a soliton equation, integrating the above equation with variation
of $\theta (x)$, we get 
\begin{equation}
\frac{d\theta }{dx}=\pm \sqrt{2\alpha Z\left( c-\cos \theta \right) }
\end{equation}%
where $c$ is a integration constant and it satisfies 
\begin{equation}
\frac{2\sqrt{2}}{\sqrt{c+1}}K\left( \frac{\sqrt{2}}{\sqrt{c+1}}\right) =%
\frac{2\pi \sqrt{\alpha Z}}{\lambda }\text{ }
\end{equation}%
according to the boundary condition $\int_{0}^{L}\partial _{x}\phi dx=0$.
Here $K\left( \frac{\sqrt{2}}{\sqrt{c+1}}\right) $ is the first class
elliptical integral function. The renormalization mass equation is
simplified as 
\begin{equation}
\mu ^{2}=\alpha Z\left[ \frac{2\sqrt{2}c}{\sqrt{c+1}}K\left( \sqrt{\frac{2}{%
c+1}}\right) -2\sqrt{2\left( c+1\right) }E\left( \sqrt{\frac{2}{c+1}}\right) %
\right] \text{ .}
\end{equation}%
$E\left( \sqrt{\frac{2}{c+1}}\right) $ is the second class elliptical
integral function. The approximate energy density of the ground state is 
\begin{equation}
\varepsilon _{0}(\lambda )=\frac{\Lambda ^{2}\sqrt{1+(\mu \Lambda ^{-1})^{2}}%
}{4\pi }+\frac{1}{\beta ^{2}}\left[ \alpha Zc-2\mu ^{2}-\frac{1}{2}\lambda
^{2}\right] \text{ .}
\end{equation}

\section{\bf \ Phase diagram\ and soliton evolvement of the ground state}

Since Eq.(11) involves elliptical function, it is difficult to be solved
explicitly. Before solving it we give some analytic discussion at special
conditions for $c$ $=1$ or $\infty $.

\subsection{Two special cases}

i) When $c=1$, we know $\lambda \Lambda ^{-1}=0$. The non-trivial solution
of Eq.(11) represents one soliton, 
\[
\cos 2\theta =\pm \tanh \sqrt{\alpha Z}x\text{ .} 
\]
For this case Nakano and Fukuyama have given a detailed discussion.$^{\cite%
{Nakano}}$ What we should note is the modification of excitation gap near
this case. After expanding the elliptical functions to linear terms of $%
\lambda \Lambda ^{-1},$ we get a expression of the renormalization mass $\mu 
$ as

\begin{equation}
(\mu \Lambda ^{-1})^{2}=(\mu _{0}\Lambda ^{-1})^{2}-\frac{2\lambda \mu
_{0}\Lambda ^{-2}}{\pi }\text{ .}
\end{equation}
$\mu _{0}\Lambda ^{-1}$ is the renormalization mass with $\lambda \Lambda
^{-1}=0$, it satisfies

\begin{equation}
(\mu _{0}{}\Lambda ^{-1})^{2}=\alpha \Lambda ^{-2}\left( \frac{\mu
_{0}\Lambda ^{-1}}{1+\sqrt{1+(\mu _{0}\Lambda ^{-1})^{2}}}\right) ^{\frac{%
\beta ^{2}}{4\pi }}\text{ .}  \label{11}
\end{equation}
If $\beta ^{2}<8\pi $, the above equation has a non-trivial solution with
nonzero value of $\mu _{0}{}\Lambda ^{-1}$. Eq.(15) implies that the
excitation gap will trail off when $\lambda \Lambda ^{-1}$ shifts from zero.
It should be noted that the ground state at finite $\lambda \Lambda ^{-1}$,
whatever small, differs from the ground state at $\lambda \Lambda ^{-1}=0$,
since the ground state is spatially inhomogenious with soliton lattice
structure in the first case, whereas it is spatially homogenous commensurate
phase in the second condition.

ii) When $c\rightarrow \infty $, we get 
\begin{equation}
\frac{1}{L}\int dx\cos \theta (x)=\frac{\alpha \Lambda ^{-2}Z}{2(\lambda
\Lambda ^{-1})^{2}}\text{ ,}
\end{equation}%
so the renormalization mass equation is simplized as 
\begin{equation}
\left( \mu \Lambda ^{-1}\right) ^{2}=(\frac{\alpha \Lambda ^{-2}}{\sqrt{2}%
\lambda \Lambda ^{-1}})^{2}\left( \frac{\mu \Lambda ^{-1}}{1+\sqrt{1+\left(
\mu \Lambda ^{-1}\right) ^{2}}}\right) ^{\frac{\beta ^{2}}{2\pi }}\text{ .}
\end{equation}

In the case of $\beta ^{2}<4\pi $, Eq.(18) always has a nonzero solution for
the renormalization mass $\mu \Lambda ^{-1}$. When $\beta ^{2}>4\pi $,
existence of a nonzero solution depends on the competition between $\alpha
\Lambda ^{-2}$ and $\lambda \Lambda ^{-1}$. In the special case of $\beta
^{2}=4\pi $, it has only zero solution when $\alpha \Lambda ^{-2}<2\sqrt{2}%
\lambda \Lambda ^{-1}$ , while it has nonzero solution when $\alpha \Lambda
^{-2}>2\sqrt{2}\lambda \Lambda ^{-1}$. For this special case the model can
be solved exactly by mapping it into a modified Thirring model, which
predicts similar result that the finite mass appears for sufficiently large $%
\alpha \Lambda ^{-2}$.

\subsection{The phase diagram of the model}

In order to give an understandable phase diagram, we numerically solve the
renormalization mass equation and soliton equation for a pair of parameters (%
$\alpha \Lambda ^{-2},\beta ^{2})$. The families of curves with constant
renormalization masses for different values of $\lambda \Lambda ^{-1}$ are
depicted in the parameter plane, see Fig.1.

With any finite $\lambda \Lambda ^{-1}$, we find that the ground state is
always in the soliton lattice phase with finite renormalization mass in the
range of $\beta ^{2}<4\pi $. The IC phase with vanishing gap only appears in
the range of $\beta ^{2}>4\pi $, increasing of $\lambda \Lambda ^{-1}$ leads
to enlarging of IC phase area. It is clear that the transition is a
continuous transition on the vertical line $\beta _{c}^{2}=4\pi $ for $%
\alpha \Lambda ^{-2}<2\sqrt{2}\lambda \Lambda ^{-1}$ since the
renormalization mass tends to zero from the left side of it. The soliton
lattice phase with finite renormalization mass is allowed in the range of $%
\beta ^{2}>4\pi $. The phase diagram is shown in Fig.2 where the transition
boundary consists of the envelope of the family $\mu \Lambda ^{-1}=$constant
with different values of $\lambda \Lambda ^{-1}$. Under the boundary the
renormalization mass vanishes, where the Hamiltonian (1) reduces to that of
a free massless boson field, whereas upon the boundary the renormalization
mass is nonzero. The first order transition occurs along this boundary in
the range of $\beta ^{2}>4\pi $.

In the area where the renormalization mass is finite, a soliton lattice
phase emerges in the real space in the ground state. The local structure of
the ground state are shown in Fig.3. It can be seen that the amplitudes of
the oscillating part of the local solitons become smaller and the periods
become shorter with increasing $\lambda \Lambda ^{-1}$. In the soliton
lattice phase, the real space in the ground state will split into many
periodic domains $\frac{2\pi (n-1)}{\lambda }<x<\frac{2\pi (n+1)}{\lambda }$
($n$ integer). In the area where the renormalization mass is zero, the
amplitude of solitons reduces to zero, the soliton lattice phase disappears,
the ground state enters into a free boson regime.

\section{\bf Conclusion}

In this paper we have studied the behavior of the ground state of a
sine-Gordon model with linear spatial modulation. The main task is to
calculate the ground state and low excitations for constant modulation
parameter. We have applied a variational procedure with respect to a
Gaussian wave functional to find an approximate ground state. Our result
shows that for $\beta ^{2}>4\pi $ IC phase grows with increasing $\lambda $,
where the low energy physics can be described by a free massless boson field
theory, while upon the boundary the excitation gap is still finite, and the
real space is no more homogeneous, a soliton lattice state appears. On the
other hand, for $\beta ^{2}<4\pi ,$ only the soliton lattice phase can
exist. With increasing spatial modulation parameter, the soliton periodic
length decreases, while the correlation length (inverse of the gap)
increases. The transition type between soliton lattice phase and IC phase
maybe of the continuous or first order, which relies on the varying of
parameter $\beta ^{2}$.

Applying above conclusion to the CDW degree freedom of the 1D extended
Hubbard model whose parameter $\beta ^{2}=8\pi \sqrt{\frac{2\pi t-2V}{2\pi
t+U+5V}}^{\cite{Voit}}$, we find the following assertion. The ground state
is always in the soliton lattice phase in the case of $U+13V>6\pi t$, the
umklapp term can not be neglected for any filling case. On the other hand
when $U+13V<6\pi t$ the ground state maybe tuned to a free massless boson
state when the band filling shifts away half filling sufficiently, only in
this case the umklapp term can be directly cancelled.

\begin{center}
\bigskip

\bigskip

{\bf Captions}
\end{center}

FIG.1 The families of curves with constant renormalization masses in the
parameter $\alpha \Lambda ^{-2}-\beta ^{2}$ plane with (a) $\lambda \Lambda
^{-1}=10^{-6}$ ,(b) $\lambda \Lambda ^{-1}=10^{-3}$ and (c) $\lambda \Lambda
^{-1}=10^{-1}$. The renormalization mass is chosen as $\mu _{1}\Lambda
^{-1}=1$, $\mu _{2}\Lambda ^{-1}=10^{-1}$, $\mu _{3}\Lambda ^{-1}=10^{-3}$, $%
\mu _{4}\Lambda ^{-1}=10^{-5}$, $\mu _{5}\Lambda ^{-1}=10^{-7}$, $\mu
_{6}\Lambda ^{-1}=10^{-9}$ and $\mu _{7}\Lambda ^{-1}=10^{-11}$.

FIG.2 Phase diagram of the model in the parameter $\alpha \Lambda
^{-2}-\beta ^{2}$ plane with different values of $\lambda \Lambda ^{-1}$.
The regions I and II represent soliton lattice phase and IC phase
respectively.\bigskip

FIG.3 The local forms of the soliton lattice phase for $\lambda \Lambda
^{-1}=0.004$ (a) and $\lambda \Lambda ^{-1}=0.01$ with $\alpha \Lambda
^{-2}=0.2$ and $\beta ^{2}=5\pi $. Here, we take$\frac{\partial \theta (x)}{%
\partial x}$ as function of coordinate $x$, since it is proportional to the
effective cosine potential.

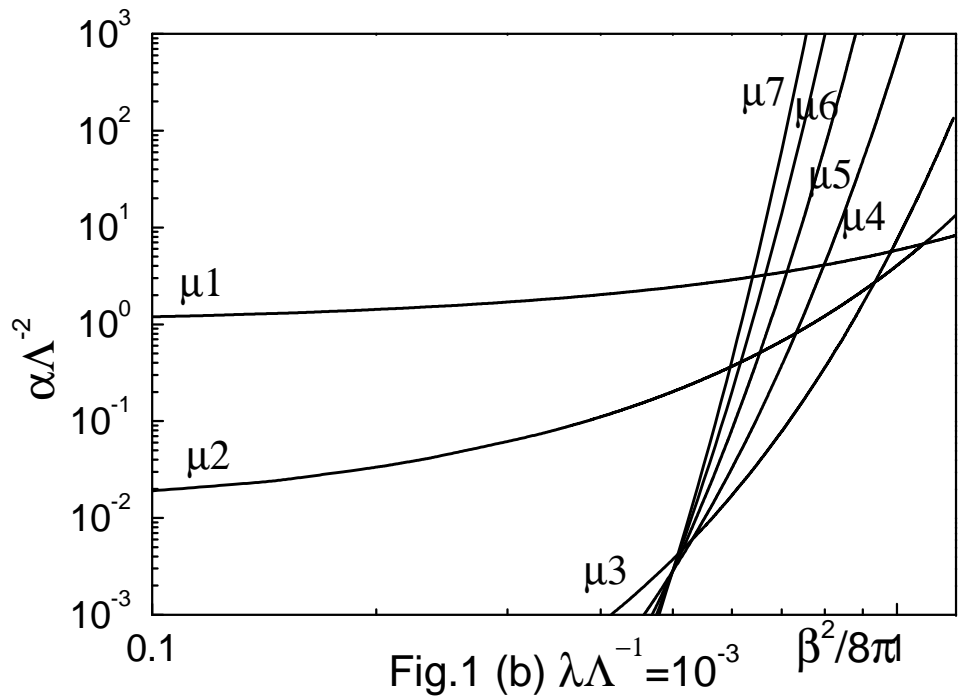
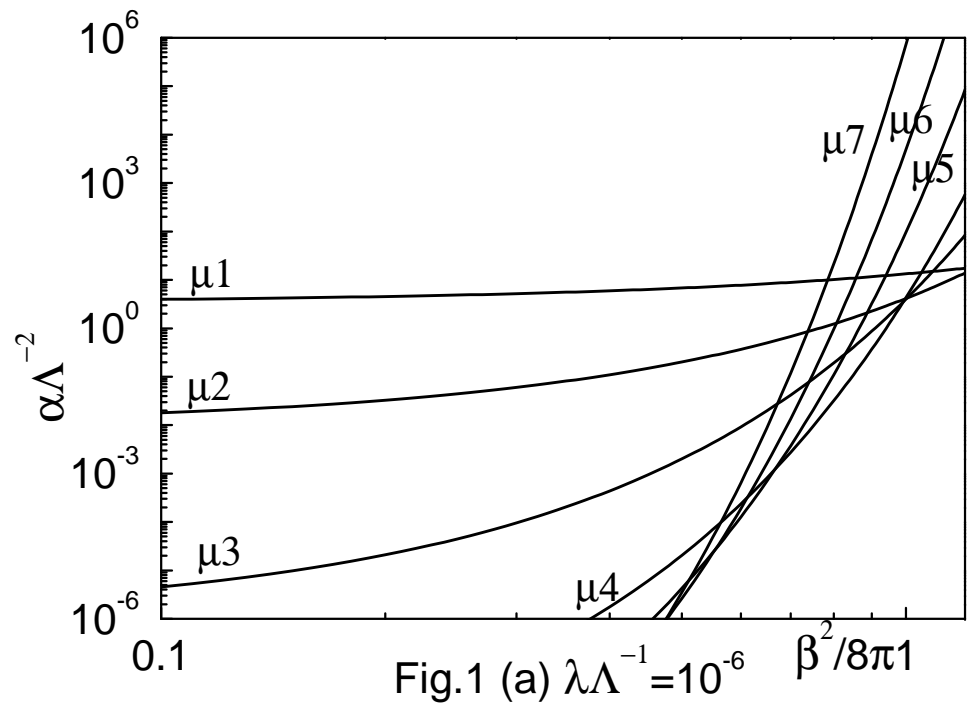
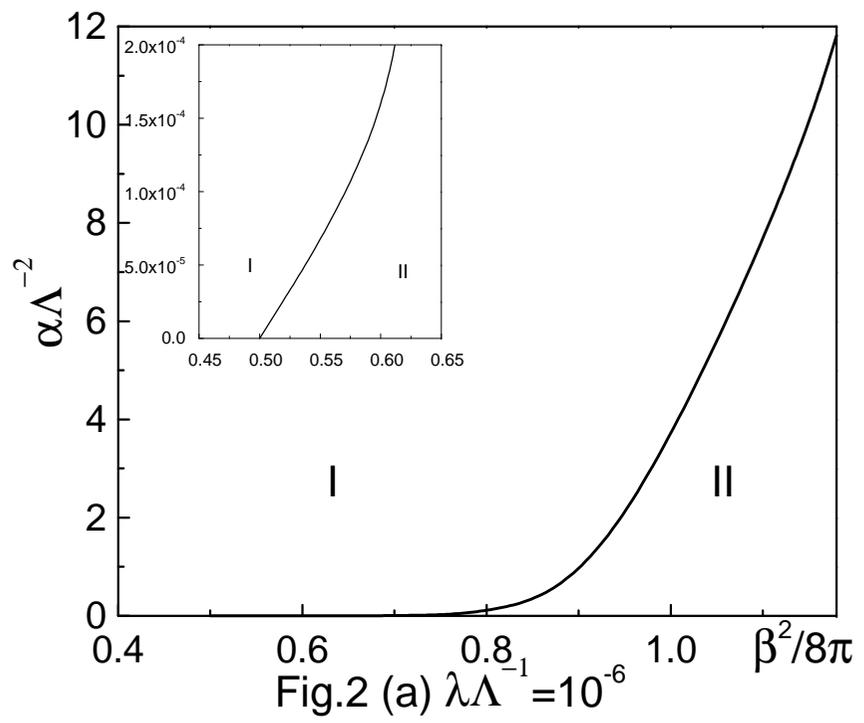
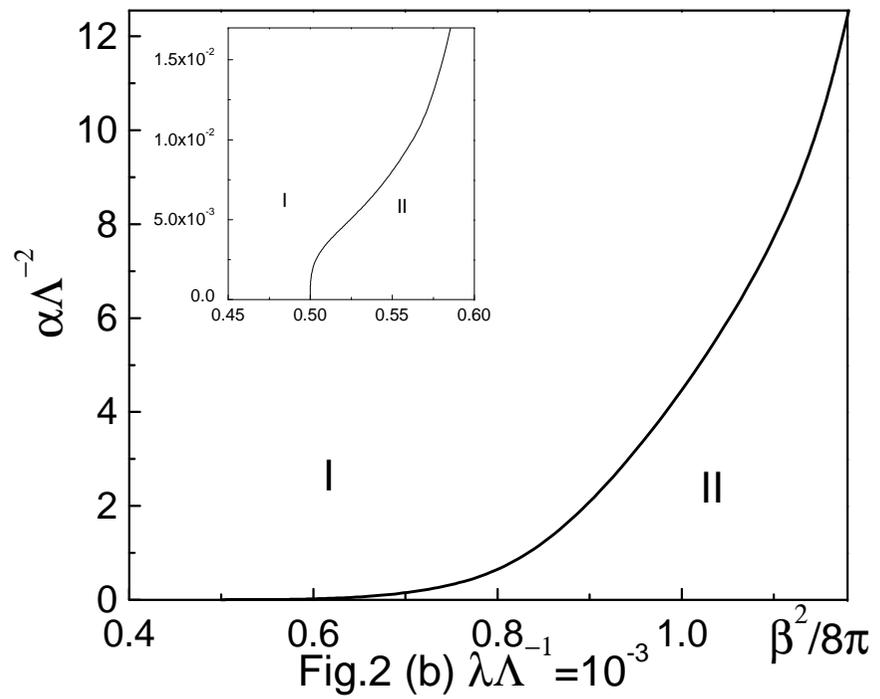
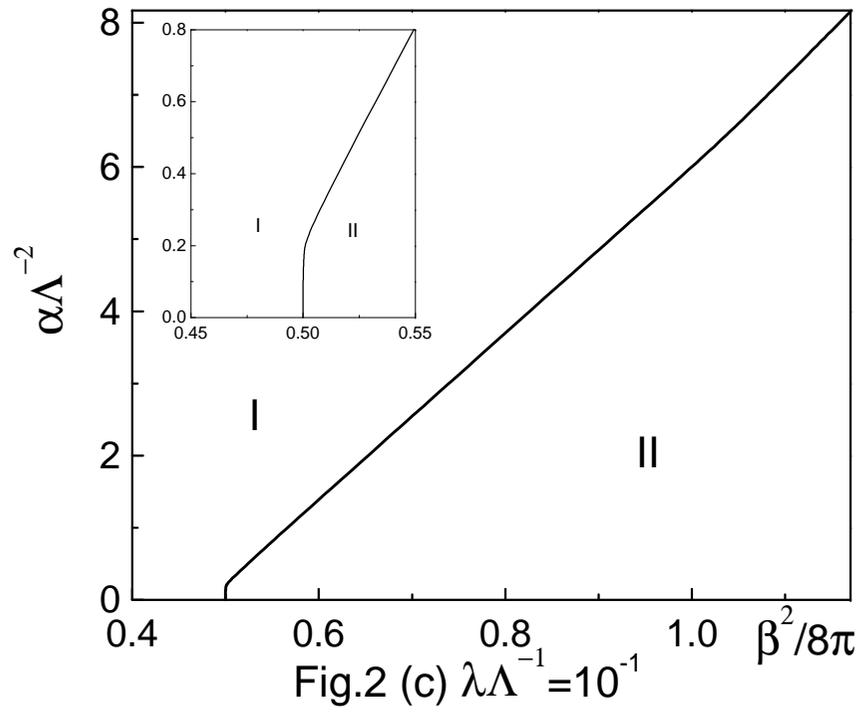
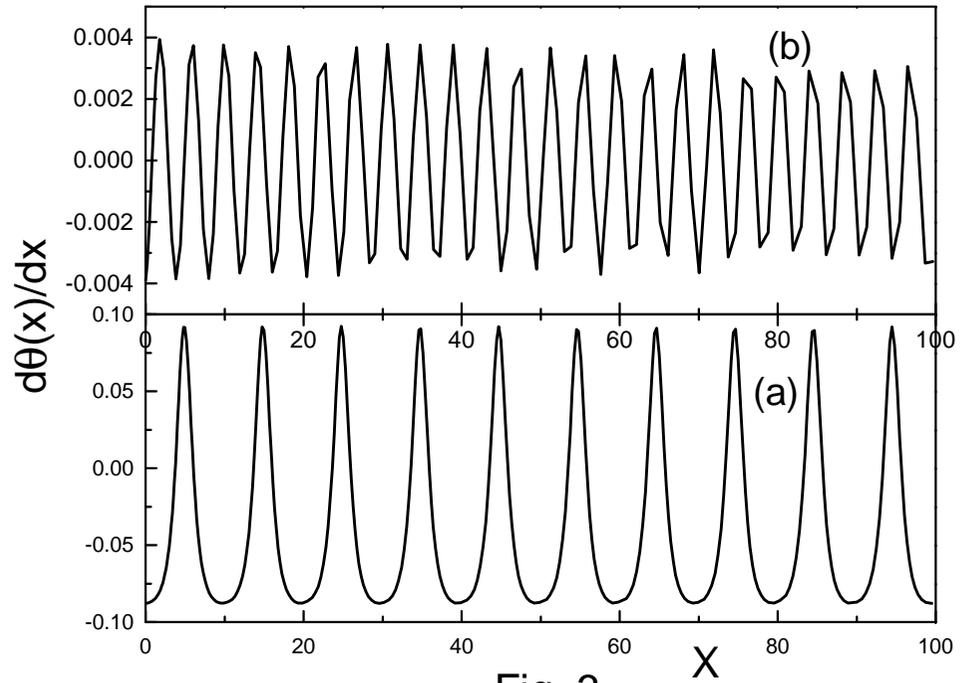
\end{document}